\documentclass[smallextended]{svjour3arx}
\smartqed  
\usepackage{graphicx}
\usepackage{hyperref}
 \journalname{Gen Relativ Gravit}
\begin{document}


\title{Comparison of particle properties in Kerr metric and in rotating
coordinates}

\titlerunning{Comparison of particle properties in Kerr metric and in rotating
coordinates}

\author{Andrey A. Grib$^{1,2}$        \and
        Yuri V. Pavlov$^{3,4}$ 
}

\authorrunning{A.A. Grib \and Yu.V. Pavlov}

\institute{A. A. Grib \at
              \email{andrei\_grib@mail.ru}
           \and
           Yu. V. Pavlov \at
              \email{yuri.pavlov@mail.ru}
           \and
${}^{1}$A.\,Friedmann Laboratory for Theoretical Physics,
St.\,Petersburg, Russia;
           \\ \\
${}^{2}$\,Theoretical Physics and Astronomy Department,
The Herzen University,\\
Moika 48, St. Petersburg 191186, Russia;
           \\ \\
${}^{3}$\,Institute of Problems in Mechanical Engineering,
Russian Academy of Sciences,\\
St. Petersburg, Russia
           \\ \\
  ${}^{4}$\,N.I.\,Lobachevsky Institute of Mathematics and Mechanics,
    Kazan Federal University, Kazan, Russia
}

\maketitle

\begin{abstract}
Properties of particles in Kerr metric are compared with properties
of particles in rotating coordinates in Minkowski space-time.
    It is shown that particles with negative and zero energies existing in
the ergosphere of the rotating black hole also exist in the region out of
the static limit in rotating coordinates in Minkowski space-time.
    Some similarities like the Penrose process and differences in both
cases are analyzed.
\end{abstract}

\section{Introduction}
\label{intro}

    It occurred that immediately after discovery of the special relativity
theory relativistic effects due to rotation were studied~\cite{Ehrenfest1909}.
    These effects are still actively discussed in
literature~\cite{EhrenfestVrash}.
    The importance of the rotating coordinate system is evident due to
the daily rotation of the Earth.

    In this paper we study the problem of the existence of particles with
negative energies in rotating coordinates.
    Here one must discriminate between rotating coordinates from rotating
reference frame.
    As it is well known rotating reference frame can be defined up to some
limit which we shall call in analogy to rotating black holes the static limit.
    The rotating coordinates can be defined everywhere.

    We show that properties of particles in rotating coordinates in Minkowski
space-time in many respects are similar to properties of particles in
the field of the rotating black hole.
    The problem of the sign of the energy in relativistic case has principal
meaning and can lead to different physical phenomena for example to
the Penrose effect in rotating black holes.
    In nonrelativistic case the potential energy of the particle is defined
up to the additive constant and depending on its definition one can have
different classification of motion of particles with negative energies.
    For example if the energy of the particle at rest on space infinity in
nonrelativistic case is taken to be zero then the sign of the sum of
the potential and kinetic energies in case of the Kepler problem defines
the bounded and unbounded orbits.
    However taking it as in relativistic case to be $mc^2$
for particle with mass $m$ ($c$ is the velocity of light) one obtains for
the full negative energy of the nonrelativistic particle moving with the
velocity $v$ on the distance $r$ from the attracting massive body with
the mass $M$
    \begin{equation}    \label{nr1}
m c^2 + \frac{m v^2}{2} - G \frac{mM}{r} <0 \ \ \Rightarrow \ \
r < \frac{G M}{c^2} = \frac{r_g}{2},
\end{equation}
    where $G$ is the Newton gravitational constant,
$r_g$ is the gravitational radius.
    So negative energy in this case can be only on distances smaller than
the gravitational radius.
    However here we shall show that negative (and zero) particle energies
are possible not only in strong gravitational fields but in the case of
absence of the gravitational field in rotating coordinates.

\section{Rotating coordinate system}
\label{sec1}

    Introduce following~\cite{LL_II}, \S\,89, the notations $r'$,
$\varphi'$, $z'$ for cylindrical coordinates and time $t$.
    The interval in Minkowski space-time in these coordinates is
    \begin{equation}    \label{v1}
d s^2 = c^2 dt^2 - d r^{\prime\,2} - r^{\prime\,2} d \varphi^{\prime\,2}
- d z^{\prime\,2}
\end{equation}
    Rotating cylindrical coordinates define as $r$, $\varphi$, $z$.
    Let the rotation axis coincides with axes $z$ and $z'$,
    \begin{equation}    \label{v2}
r'=r, \ \ \ \ z'=z, \ \ \ \ \varphi' = \varphi - \Omega t,
\end{equation}
    where $\Omega \ge 0$ is the angular velocity of rotation.
    Putting~(\ref{v2}) into~(\ref{v1}) one obtains the expression for
the interval in rotating coordinate system
    \begin{equation}    \label{v3}
d s^2 = (c^2 - \Omega^2 r^2)\, dt^2 + 2 \Omega r^2 d \varphi\, d t -
d r^{2} - r^{2} d \varphi^{2} - d z^{2},
\end{equation}
    \begin{equation}    \label{vgvik}
\left( g_{ik} \right) =
\left( \begin{array}{crcr}
\displaystyle 1 - \frac{\Omega^2 r^2}{c^2} & 0
& \displaystyle \ \frac{\Omega r^2}{c} & 0 \\[2mm]
0 & - 1 & 0 & 0 \\[2mm]
\displaystyle  \frac{\Omega r^2}{c} & 0 &
- r^2 & 0 \\[2mm]
0 & 0 & 0 & -1
\end{array} \right), \ \
\left( g^{ik} \right) =
\left( \begin{array}{rrcr}
1 & 0 & \displaystyle \frac{\Omega}{c} & 0 \\[2mm]
0 & - 1 & 0 & 0 \\[2mm]
\displaystyle  \frac{\Omega}{c} & 0 &
\displaystyle \ \frac{\Omega^2}{c^2} - \frac{1}{r^2} & 0 \\[2mm]
0 & 0 & 0 & -1
\end{array}
\right),
\end{equation}
    $ i,k = 0,1,2,3 $.

    In book~\cite{LL_II} it is said that
``the rotating system of reference can be used only out to distances equal
to $ c / \Omega$.
    In fact, from~(\ref{v3}) we see that for $ r > c / \Omega$,
$g_{00}$ becomes negative, which is not admissible.
    The inapplicability of the rotating reference system at large distances
is related to the fact that there the velocity would become greater than
the velocity of light, and therefore such a system cannot be made up from
real bodies.''
    The same opinion is present in the book~\cite{Fok} where going to
rotation reference frame is considered to be possible only for
distances $ r < c / \Omega$.

    Note however that in spite of $g_{00}=0$ for $ r= c / \Omega$
the metric~(\ref{v3}) remains nondegenerate:
$ {\rm det}\! \left( g_{ik} \right) = - r^2$.

    If one uses the rotating coordinate system for the Earth then the angular
velocity is $\Omega_\oplus \approx 7.29 \cdot 10^{-5}$\,s, and the distance
where $g_{00}$ is zero is $ c/\Omega_\oplus = 4.11 \cdot 10^9$\,km.
    This distance is smaller than the orbit of the planet
Neptune $ r \approx 4.5\cdot 10^9$\,km, but farther than the orbit of
Uranus $ r \approx 2.9\cdot 10^9$\,km.
    It is evident that the rotating Earth coordinate system must be used not
only for Neptune but much farther.
    Surely one cannot realize the ``reference frame'' by some immovable rods
or by any bodies at rest in this system.
    Finite value of the light velocity surely prohibits to do it on some
large distances from the rotation axis.
    However due to general covariance of general relativity one can
use ``the rotating coordinate system''.
    So further in this paper we shall study motion of particles in
rotating coordinate system.

    There can be no physical body in rotating coordinate system being at rest
on distances $ r > c / \Omega$ due to impossibility of motion with the
velocity larger than that of light.
    So the surface $ r= c / \Omega$ plays the role of the static limit well
known in Kerr metric for the rotating black hole.
    We shall show in this paper that out of this surface, $ r > c / \Omega$,
the particle energy in rotating coordinate system can have negative value
as it is the case for the Kerr metric.
    So let us write the formulas for the Kerr metric and compare them with
those in rotating coordinates in Minkowski space-time.

\section{The Kerr metric of the rotating black hole}
\label{secKerrVr}

    The interval in Kerr metric of the rotating black hole~\cite{Kerr63}
in Boyer-Lindquist coordinates~\cite{BoyerLindquist67} is
    \begin{equation}
d s^2 = \frac{\rho^2 \Delta}{\Sigma^2}\,c^2 d t^2 -
\frac{\sin^2\! \theta}{\rho^2} \Sigma^2 \, ( d \varphi - \Omega d t)^2
\label{Kerr}
- \frac{ \rho^2}{\Delta}\, d r^2 - \rho^2 d \theta^2 ,
\end{equation}
    where
    \begin{equation} \label{Delta}
\rho^2 = r^2 + \frac{a^2}{c^2} \cos^2 \! \theta, \ \ \ \ \
\Delta = r^2 - \frac{2 G M r}{c^2} + \frac{a^2}{c^2},
\end{equation}
    \begin{equation} \label{Sigma}
\Sigma^2 = \left( r^2 + \frac{a^2}{c^2} \right)^2 -
\frac{a^2}{c^2} \sin^2\! \theta\, \Delta , \ \ \ \
\Omega = \frac{2 G M r a}{\Sigma^2 c^2} ,
\end{equation}
    $M$ is the mass of the black hole, $ aM $ is angular momentum.
We suppose ${0 \le a \le G M/c }$.
    The covariant components of metric tensor in the coordinates
$ x^0 = c t$, $x^1 = r$, $x^2= \theta$, $x^3= \varphi $ are
    \begin{equation}
\left( g_{ik} \right) =\left(
\begin{array}{cccc}  \displaystyle
\frac{S}{\rho^2} & 0 & 0 &
\displaystyle \frac{2 G M r a \sin^2\! \theta}{\rho^2 c^3} \\[2mm]
0 &  \displaystyle -\frac{\rho^2}{\Delta} & 0 & 0 \\[2mm]
0 & 0 &  -\rho^2 & 0 \\[2mm]
\displaystyle \frac{2 G M r a \sin^2\! \theta}{\rho^2 c^3} & 0 & 0 &
\displaystyle - \frac{\sin^2 \! \theta }{ \rho^2 } \, \Sigma^2
\end{array}
\right),
\label{gKov}
\end{equation}
    \begin{equation} \label{S}
S (r, \theta) = r^2 - \frac{2 G M r}{c^2} + \frac{a^2}{c^2} \cos^2 \! \theta.
\end{equation}
    The event horizon for the Kerr black hole is given by
    \begin{equation}
r = r_H \equiv \frac{G M}{c^2} + \sqrt{\frac{G^2M^2}{c^4} - \frac{a^2}{c^2}} ,
\label{Hor}
\end{equation}
    $\Delta (r_H) =0$.
    $\Delta (r) >0$ outside of the horizon of events.
    The surface
    \begin{equation}
r = r_C \equiv \frac{G M}{c^2} - \sqrt{\frac{G^2M^2}{c^4} - \frac{a^2}{c^2}} ,
\label{HorC}
\end{equation}
    is the Cauchy horizon.
    The surface of the static limit is defined by
    \begin{equation}
r = r_1 \equiv \frac{GM}{c^2} + \sqrt{\frac{G^2 M^2}{c^4} -
\frac{a^2}{c^2} \cos^2 \theta} .
\label{Lst}
\end{equation}
    It is obvious that $ S (r_1, \theta)=0$.
    The region of space-time between the static limit and the event horizon
is called ergosphere~\cite{MTW}, \cite{NovikovFrolov}.
    Inside the ergosphere one has $ S (r, \theta) < 0$.
    Note that
    \begin{equation} \label{SigmaDr}
\Sigma^2 = \left( r^2 + \frac{a^2}{c^2} \right) \rho^2 +
\frac{2 r G M a^2}{c^4} \sin^2\! \theta > 0, \ \ {\rm if} \ \ r > 0.
\end{equation}
    Using the relation
    \begin{equation} \label{SSrho}
S\, \Sigma^2 + \frac{4 G^2 M^2 r^2 a^2}{c^6} \sin^2\! \theta = \rho^4 \Delta ,
\end{equation}
    one can find
$ {\rm det}\! \left( g_{ik} \right) = - \rho^4 \sin^2\! \theta$.
    That is why for $\rho \ne 0$, $\theta \ne 0,\pi$,
the metric~(\ref{Kerr}) is nondegenerate.
    Further for Kerr metric we will use the system of units in
which $G=c=1$.

    Using the relation~(\ref{SSrho}) one can write the equations of geodesics
for the Kerr metric~(\ref{Kerr})
(see~\cite{Chandrasekhar}, Sec.~62 or~\cite{NovikovFrolov}, Sec.~3.4.1) as
    \begin{equation} \label{geodKerr1}
\rho^2 \frac{d t}{d \lambda } = \frac{1}{\Delta}
\left( \Sigma^2 E - 2 M r a J \right), \ \ \
\rho^2 \frac{d \varphi}{d \lambda } = \frac{1}{\Delta}
\left( 2 M r a E + \frac{S J}{\sin^2\! \theta} \right),
\end{equation}
    \begin{equation} \label{geodKerr3}
\rho^2 \frac{d r}{d \lambda} = \sigma_r \sqrt{R}, \ \ \ \ \
\rho^2 \frac{d \theta}{d \lambda} =\sigma_\theta \sqrt{\Theta},
\end{equation}
    \begin{equation} \label{geodR}
R = \Sigma^2 E^2 - \frac{S J^2}{\sin^2 \theta } - 4 M r a E J -
\Delta \left[ m^2 \rho^2 + \Theta \right],
\end{equation}
    \begin{equation} \label{geodTh}
\Theta = Q - \cos^2 \! \theta \left[ a^2 ( m^2 - E^2) +
\frac{J^2}{\sin^2 \! \theta} \right].
\end{equation}
    Here $E={\rm const}$ is the energy (relative to infinity) of the moving
particle,
$J$ is the conserved angular momentum projection on the rotation axis,
$m$ is the rest mass of the particle,
$\lambda $ --- the affine parameter along the geodesic.
    For the particle with $m \ne 0$ the parameter $\lambda = \tau /m$, where
$\tau$ is the proper time.
$Q$ is the Carter constant.
    $Q=0$ for the motion in the equatorial plane $(\theta = \pi/2)$.
    The constants $\sigma_r, \sigma_\theta = \pm 1$ define the direction
of motion in coordinates $r, \theta$.

    From~(\ref{geodKerr3}) follows that the parameters characterizing any
geodesic must satisfy the conditions
    \begin{equation} \label{ThB0}
R \ge 0, \ \ \ \ \ \Theta \ge 0 .
\end{equation}
    For the geodesic being the trajectory of the test particle moving outside
the event horizon one must have the condition of motion ``forward in time''
    \begin{equation} \label{ThB0t}
d t / d \lambda > 0 .
\end{equation}
    The conditions~(\ref{ThB0}), (\ref{ThB0t}) lead to inequalities for
possible values of the energy $ E $ and angular momentum projection $J$ of
the test particle at the point with coordinates $(r, \theta)$ with fixed
value~$\Theta \ge 0$~\cite{GribPavlov2013}.

    Outside the ergosphere $ S(r, \theta) >0 $,
    \begin{equation} \label{EvErg}
E \ge \frac{1}{\rho^2} \sqrt{(m^2 \rho^2 + \Theta) S},
\ \  \ J \in \left[ J_{-} (r,\theta), \ J_{+} (r,\theta) \right],
\end{equation}
    \begin{equation}
J_{\pm} (r,\theta) = \frac{\sin \theta}{S} \left[ - 2 r M a E \sin \theta \pm
\sqrt{ \Delta \left( \rho^4 E^2 - (m^2 \rho^2 + \Theta) S \right)} \right].
\label{Jpm}
\end{equation}

    On the boundary of the ergosphere (for $\theta \ne 0, \pi$)
    \begin{equation} \label{JgErg}
r = r_1(\theta) \  \Rightarrow \  E \ge 0, \ \
J \le E \left[ \frac{M r_1(\theta) }{a} + a \sin^2 \! \theta \left(
1 - \frac{m^2}{2 E^2} - \frac{\Theta}{4 M r_1(\theta) E^2} \right) \right].
\end{equation}
    The value $E=0$ is possible on the boundary of the ergosphere when $m=0$
and $\Theta=0$.
    In this case one can have any value of $J <0$.

    Inside the ergosphere $ r_H < r < r_1(\theta) $, $S < 0$
    \begin{equation} \label{lHmdd}
J \le \frac{\sin \theta}{- S} \left[ 2 r M a E \sin \theta -
\sqrt{ \Delta \left( \rho^4 E^2 - (m^2 \rho^2 + \Theta) S \right)} \right].
\end{equation}
    and the energy of the particle as it is known can be as positive
as negative.

    As it is seen from~(\ref{JgErg}), (\ref{lHmdd}) on the boundary and inside
the ergosphere the angular momentum projection of particles moving along
geodesics can be negative and it can be any large in absolute value number
for the fixed value of the energy.
    This property is valid not only for the Kerr
metric~\cite{GribPavlov2013,GribPavlov2012}, but in the ergosphere of
any black hole with the axially symmetric metric~\cite{Zaslavskii13c}.
    Note that in the vicinity of horizon from~(\ref{lHmdd}) one has
    \begin{equation} \label{JH}
J(r) \le J_H = \frac{2 r_H M E}{a}, \ \ \ r \to r_H.
\end{equation}

    If the the values of $J$ and $\Theta \ge 0$ are given then
from~(\ref{ThB0}), (\ref{ThB0t}) one has at any point outside the horizon
    \begin{equation}
E \ge \frac{1}{\Sigma^2} \left[ 2 M r a J + \sqrt{ \Delta \left(
\frac{ \rho^4 J^2 }{\sin^2\! \theta } + \left( m^2 \rho^2 + \Theta \right)
\Sigma^2 \right) } \right].
\label{EVnHS}
\end{equation}
    The lower boundary $E$ corresponds to $R=0$.
    As it is seen from~(\ref{EVnHS}) negative energies can be
only in case of the negative value of $J$ of the angular momentum of
the particle and for $r$ such as $2 M r a |\sin \theta| > \rho^2 \sqrt{\Delta}$,
i.e., in accordance with~(\ref{SSrho}), in the ergosphere.

\section{The energy of the point particles in curved space-time}
\label{secNewE}

    As it is known~\cite{Chandrasekhar},  equations of geodesics in
the space-time with the interval $ds^2 = g_{ik} dx^i dx^k$ can be obtained
from the Lagrangian
    \begin{equation}
L = \frac{g_{ik}}{2}\, \frac{ d x^i}{d \lambda} \frac{ d x^k}{d \lambda},
\label{Lgik}
\end{equation}
    where $\lambda$ defined earlier in~(\ref{geodKerr1}),
(\ref{geodKerr3}) is the affine parameter on the geodesic.

    Generalized momenta are by definition
    \begin{equation}
p_i  \stackrel{\rm def}{=} \frac{\partial L}{\partial \dot{x}^i}
= g_{ik} \frac{d x^k}{d \lambda } ,
\label{Lpdef}
\end{equation}
    where $ \dot{x}^i = d x^i/d \lambda $.
    It is evident that contravariant components of the generalized momenta are
equal to corresponding velocities:
    \begin{equation}
p^i = \frac{d x^i}{ d \lambda}.
\label{pKontra}
\end{equation}
    For timelike geodesic let us normalize the affine
parameter $\lambda = \tau /m$, where $\tau$ is the proper time of the moving
particle with mass $m$.
    Then
    \begin{equation}
p_i p_k  g^{ik} = m^2 c^2.
\label{pipkm2}
\end{equation}

    If the metric components $g_{ik}$ do not depend on some coordinate then
the corresponding canonical momentum (the corresponding covariant component)
is conserved in motion along the geodesic due to Euler-Lagrange equations:
    \begin{equation}
\frac{d }{d \lambda} \frac{\partial L}{\partial \dot{x}^i} -
\frac{\partial L}{\partial x^i} = 0 \ \ \ \Rightarrow \ \ \
p_i = \frac{\partial L}{\partial \dot{x}^i} = {\rm const}.
\label{LEL}
\end{equation}

    In the general case independence of the metric on some coordinate does not
lead to the conservation of the corresponding contravariant momentum component!
    For example consider free moving particle in Minkowski space-time.
    In Cartesian coordinates all components of the energy-momentum vector
as covariant as contravariant are conserved  
    \begin{equation}    \label{v6}
p'^{\, i} = \left( \frac{E'}{c}, \ {\bf p'} \right), \ \ \ \
p'_{\, i} = \left( \frac{E'}{c}, \ {\bf - p'} \right),
\end{equation}
    where ${\bf p'}$ is usual three momentum (see \cite{LL_II}, \S\,9).
    In cylindrical coordinates
    \begin{equation}    \label{v6d}
p'^{\, i} = m \frac{d x^{\prime\,i}}{d \tau} = \left( \frac{E'}{c}, \
p^{\prime\,r}, \ p^{\prime \varphi}, \ p^{\prime z} \right), \ \ \
p'_{\, i} = \left( \frac{E'}{c}, \ - p^{\prime\,r},
\ - L'_z, \ - p^{\prime z} \right),
\end{equation}
    where
    \begin{equation}    \label{v6kk}
L'_z = r^2 p^{\prime \varphi} = m r^2 \frac{d \varphi' }{d \tau}
= \frac{E'}{c^2} r^2 \frac{d \varphi' }{d t}
\end{equation}
    is the projection of the angular momentum on the axis $(OZ)$.
    The metric components do not depend on the angle $\varphi$
and the covariant momentum component $- L'_z$ is conserved.
    Contravariant component $p^{\prime \varphi}$ evidently is not conserved.

    In case when the components of metric do not depend on time $t$
the zero covariant component of the momentum $p_0$ of the freely moving
particle is conserved and it is equal to the particle energy divided
on the light velocity~$E/c$:
    \begin{equation}    \label{dE1}
E=p_0 c.
\end{equation}
    Let us show that this expression for the energy can be obtained differently
using Killing vector.

    Let the space-time have the timelike Killing vector~$\zeta^i$
orthogonal to some set of spacelike hypersurfaces~\{$\Sigma$\}.
    From the definition of the Killing vector one has
    \begin{equation} \label{NEnergy1}
\nabla^i \zeta^k + \nabla^k \zeta^i = 0 \,.
\end{equation}
    Let $T_{ik}$ be the metrical energy-momentum tensor of some field or
covariantly conserved energy-momentum tensor of some matter.
    The translational symmetry with the generator~$\zeta^i$
leads to the conservation of the quantity
    \begin{equation} \label{NEnergy2}
E^{(\zeta)} = \int_\Sigma T_{ik}\, \zeta^i \, d \sigma^k ,
\end{equation}
     which follows from the covariant conservation of~$T_{ik}$
and Eq.~(\ref{NEnergy1}) leading to~$ \nabla^i (T_{ik} \zeta^k)=0 $.
    The quantity~$ E^{(\zeta)} $ plays the role of the energy.

    Let us find $ E^{(\zeta)} $ for a classical pointlike particle with
the mass~$m$.
    By variation of the action
    \begin{equation} \label{NEnergy4}
S = - mc \int \! ds
\end{equation}
    one obtains the energy-momentum tensor of the classical pointlike particle,
located at the point with coordinates~$x_{p}$ as
    \begin{equation} \label{NEnergy5}
T^{ik}(x) = - \frac{2 c}{\sqrt{ |g|}} \, \frac{\delta S}{\delta g_{ik}}
= \frac{m c^2 }{\sqrt{|g|}} \int \! ds \, \frac{d x^i}{d s}\,
\frac{d x^k}{d s}\, \delta^4 ( x - x_{p}) .
\end{equation}
    The value of~(\ref{NEnergy2}) for a pointlike particle in a general metric
and for an arbitrary vector~$ \zeta^i $ is
    \begin{equation} \label{NEnergyEgen}
E^{(\zeta)} = m c^2\, \frac{dx^i}{ds}\,  g_{ik} \zeta^k =
m c^2 \, (u, \zeta) = c (p, \zeta),
\end{equation}
    where $ u^i= dx^i / ds $ is the four velocity, \,
$ p^i= m\, c\, dx^i / ds $ is the four momentum.
    Note that the value~(\ref{NEnergyEgen}) is conserved along the geodesic
for any Killing vector which can be not only timelike
(see Problem 10.10. in book~\cite{LPPT}).

    The value of the energy $E^{(\zeta)}$ evidently depends on the choice of
the Killing vector~$\zeta$.
    If metric does not depend on time and the Killing vector is chosen so that
$\zeta = (1,0,0,0)$, then the energy $E^{(\zeta)}$ is equal to~(\ref{dE1}).

\section{The energy and the static limit for the metric with rotation}
\label{secgVr}

    Here we obtain some general limitations on the parameters of particle
motion in metric with rotation for the interval
    \begin{equation}\label{gik}
d s^2 = g_{00} c^2 d t^2 + 2 g_{0 \varphi} c\, d t d \varphi +
g_{\varphi \varphi}\, d \varphi^2 + g_{rr}\, d r^2 +
g_{\theta \theta}\, d \theta^2
\end{equation}
    in the region where $ g_{\varphi \varphi} < 0 $, $g_{rr} < 0$,
$ g_{\theta \theta} < 0$.
    The Kerr metric~(\ref{Kerr}) out of the event horizon $(r > r_H)$
and the metric of the rotating coordinate system~(\ref{v3}),
if one changes $z \to \theta$ satisfy these conditions.

    One must have $ds^2 \ge 0$ for the interval for any points of the
world line of the moving particle and this leads to the inequality
    \begin{equation}
g_{00} c^2 d t^2 + 2 g_{0 \varphi} c\, d t\, d \varphi +
g_{\varphi \varphi}\, d \varphi^2 \ge 0.
\label{omN1}
\end{equation}
    So if $d \varphi \ne 0$ one must have
$ (g_{0 \varphi})^{2} - g_{00} g_{\varphi \varphi} > 0 $ (the sign of equality
is excluded by the condition of the nondegeneracy of metric~(\ref{gik}))
and the angular velocity $\omega = d \varphi / d t $ for any particle is in
the limits~\cite{MTW}
    \begin{equation}
\Omega_1 \le \omega \le \Omega_2, \ \ \ \
\Omega_{1,2} = \frac{c}{- g_{\varphi \varphi} } \left( g_{0 \varphi} \mp
\sqrt{ (g_{0 \varphi})^{2} - g_{00} g_{\varphi \varphi}} \, \right).
\label{omN2}
\end{equation}
    If $g_{00} \le 0 $ then the rotation is possible only in one direction.
    The coordinates corresponding to $g_{00} = 0 $ define the static limit.
    For the rotating coordinate system~(\ref{v3}) the static limit corresponds
to the value $r = c/\Omega$.
    In the region out of the event horizon of the Kerr black hole the static
limit is defined by the formula~(\ref{Lst}).

    Note that the well known in nonrelativistic mechanics relation between
the angular and linear velocities of rotation $v = \omega r$ in general
does not correspond to relative velocity of rotation of one particles
relative to other particles at rest.
    For the rotating coordinate system for $r > c / \Omega$ one could obtain
the value larger than that of the velocity of light.
    However the condition $ds^2\ge 0$ for any values of 4-coordinates of
the observable particle guarantees that its velocity relative to
any particle located in its vicinity cannot be larger than the velocity
of light.

    Let us find the expression for the energy of the particle in metric
with rotation~(\ref{gik}).
        Due to expressions of the contravariant metric components
    \begin{equation}
g^{00} = \frac{-g_{\varphi \varphi} }{
(g_{0 \varphi})^{2} \!-\! g_{00} g_{\varphi \varphi}} > 0, \
g^{\varphi \varphi} = \frac{-g_{00} }{
(g_{0 \varphi})^{2} \!-\! g_{00} g_{\varphi \varphi}}, \
g^{0 \varphi} = \frac{g_{0 \varphi} }{
(g_{0 \varphi})^{2} \!-\! g_{00} g_{\varphi \varphi}},
\label{pvik}
\end{equation}
    \begin{equation}
(g^{0 \varphi})^2 - g^{00} g^{\varphi \varphi} =
\frac{1}{(g_{0 \varphi})^{2} \!-\! g_{00} g_{\varphi \varphi} } >0, \ \ \
g^{rr} = \frac{1}{g_{rr}} < 0 , \ \ \
g^{\theta \theta} = \frac{1}{g_{\theta \theta}}<0,
\label{pvik2}
\end{equation}
    from~(\ref{pipkm2}) one obtains
    \begin{eqnarray}
p_0 &=& \frac{1}{g^{00}} \left[ - p_\varphi g^{0 \varphi}
\pm \sqrt{p_\varphi^2 ( g^{0 \varphi})^2 -
g^{00} \left( p_\varphi^2 g^{\varphi \varphi} + p_r^2 g^{rr} +
p_\theta^2 g^{\theta \theta} - m^2 c^2 \right)} \right] =
\nonumber \\
&=& p_\varphi \frac{ g_{0 \varphi}}{g_{\varphi \varphi}}
\pm \sqrt{ \frac{(g_{0 \varphi})^{2} -
g_{00} g_{\varphi \varphi}}{-g_{\varphi \varphi}}}
\sqrt{m^2 c^2 -  \frac{ p_\varphi^2}{g_{\varphi \varphi}} -
\frac{ p_r^2}{g_{rr}} - \frac{ p_\theta^2}{g_{\theta \theta}}}.
\label{pvik3}
\end{eqnarray}
    In region where $g_{00} >0$ the sign of $p_0$ coincides with the sign of
the root in the right hand side of this formula.
    Suppose that in the region with $g_{00} >0$ the energy of particles is
positive then the minus sign in~(\ref{pvik3}) must be put away.
    Then one obtains for the energy of particles
    \begin{equation}
E = c p_\varphi \frac{ g_{0 \varphi}}{g_{\varphi \varphi}} +
c \sqrt{ \frac{(g_{0 \varphi})^{2} -
g_{00} g_{\varphi \varphi}}{-g_{\varphi \varphi}}}
\sqrt{m^2 c^2 -  \frac{ p_\varphi^2}{g_{\varphi \varphi}} -
\frac{ p_r^2}{g_{rr}} - \frac{ p_\theta^2}{g_{\theta \theta}}}.
\label{pvik4}
\end{equation}

    As it is clear from~(\ref{pvik4}) the necessary and sufficient condition
of the existence of states with different signs of the energy in the same
region of coordinate values an metric with rotation~(\ref{gik}) is
the condition $g_{00} <0$.
    If $g_{00} =0$ then together with the positive sign of the energy states
with zero energy and nonzero momentum are possible so that \ $E=0$ when
$p_\varphi g_{0 \varphi} > 0$, \ $p_r$, $p_\theta$, $m=0$.

    So in metric with rotation~(\ref{gik}) the static limit coincides with
the boundary of existence of states with negative energy and is defined by
the condition $g_{00}=0$.
    Inside this region $g_{00}<0$ and no particle can be at rest in
the coordinate system~(\ref{gik}).
    Here the energies can be positive, negative and zero.

        If the particle has zero angular momentum then due to
the definition~(\ref{Lpdef})
    \begin{equation}
p_\varphi= g_{0 \varphi} \frac{c dt}{ d \lambda} +
g_{\varphi \varphi} \frac{d \varphi}{ d \lambda} = 0
\label{E0mi}
\end{equation}
    so that the angular velocity is
    \begin{equation}
p_\varphi=0 \ \ \ \Rightarrow \ \ \ \frac{d \varphi}{ d t} =
- \frac{c\, g_{0 \varphi} }{g_{\varphi\varphi} }.
\label{E0miom}
\end{equation}
    For the metrics~(\ref{v3}), (\ref{Kerr}) this angular velocity is equal
to the corresponding values~$\Omega$.

    If the particle has zero energy then due to the definition~(\ref{Lpdef})
    \begin{equation}
g_{00} \frac{c dt}{ d \lambda} + g_{0 \varphi} \frac{d \varphi}{ d \lambda} = 0
\label{E0gp}
\end{equation}
    and its angular velocity is
    \begin{equation}
E=0 \ \ \ \Rightarrow \ \ \
\frac{d \varphi}{ d t} = - \frac{c\, g_{00} }{g_{0 \varphi} } = \Omega_{0}.
\label{E0phig}
\end{equation}
    The angular velocity of any particle with zero energy does not depend on
the value of mass and angular momentum of the particle and is defined
by the formula~(\ref{E0phig}).

    The values of angular velocities $\omega$ for particles with positive
and negative energies are in intervals $[ \Omega_0, \Omega_2]$ and
$[ \Omega_1, \Omega_0]$ correspondingly if $ g_{0 \varphi} >0$,
and in intervals $[ \Omega_1, \Omega_0]$, $[ \Omega_0, \Omega_2]$,
correspondingly if $ g_{0 \varphi} < 0 $.

\begin{figure}[th]
\centering
  \includegraphics[width=61mm]{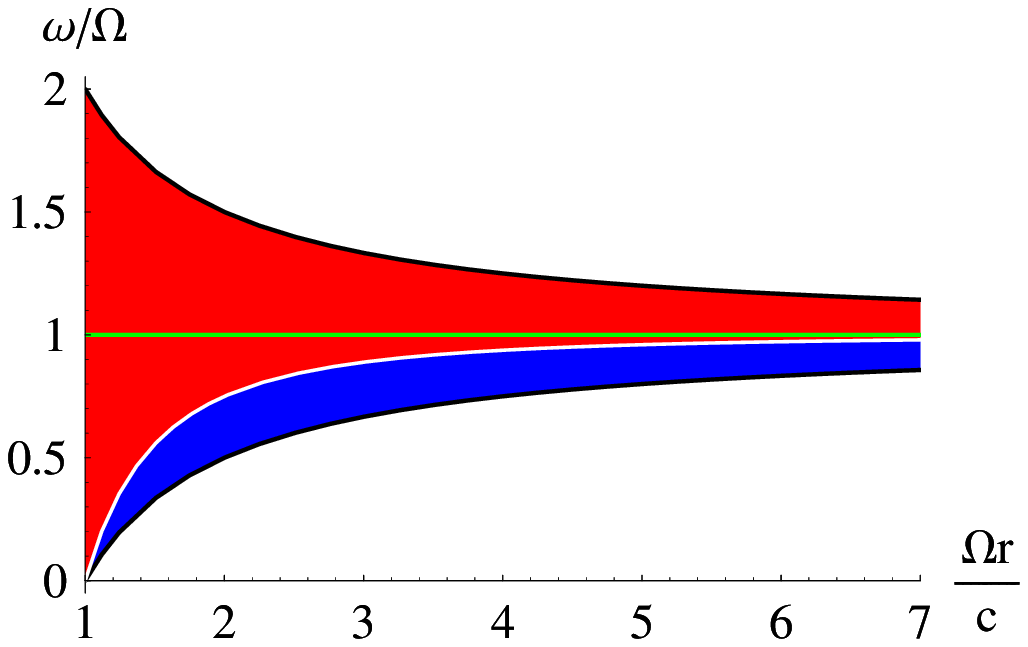}
  \includegraphics[width=55mm]{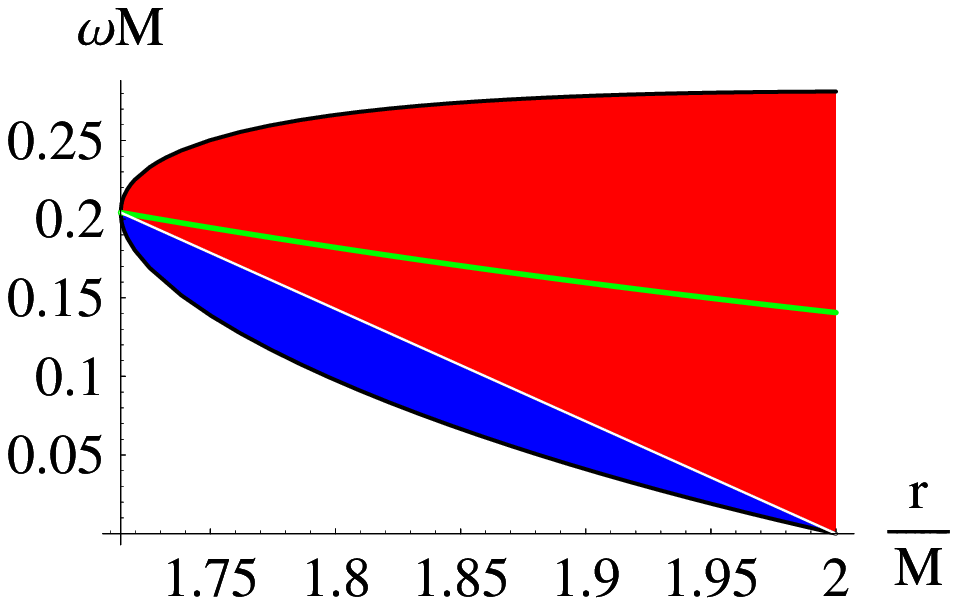}
\caption{Possible angular velocities in rotating coordinate system (left)
and in ergosphere of Kerr metric for $\theta = \pi/2$ and $a/M=0.7$ (right).
The red region --- angular velocities of particles with positive
energy, the blue region --- for particles with negative energy,
the white line --- $E=0$, the green line --- $p_\varphi=0$.}
\label{VrOmega}
\end{figure}

    For the metric of uniformly rotating coordinate system~(\ref{v3})
the values $\Omega_{0,1,2}$ (see (\ref{omN2}), (\ref{E0phig})) are
    \begin{equation}
\Omega_0 = \Omega \left( 1 - \frac{c^2}{\Omega^2 r^2} \right),
\ \ \ \ \Omega_{1,2} = \Omega \mp \frac{c}{r}.
\label{Om012Vr}
\end{equation}
    For Kerr metric~(\ref{Kerr})
    \begin{equation}
\Omega_0  = \frac{- S(r)}{2 r M a \sin^2 \! \theta}, \ \ \
\Omega_{1,2} = \Omega \mp \frac{ \rho^2 \sqrt{\Delta}}{
\sin \theta \, \Sigma^2}.
\label{OmK012}
\end{equation}
    Possible values of the angular velocity of particles with positive,
negative and zero energies are shown on Fig.~\ref{VrOmega}.

    Note that for the Kerr metric
    \begin{equation}
\Omega_{(0,1,2)} \to \omega_{Bh} = \frac{a}{2 M r_H}, \ \ r \to r_H.
\label{OmBh}
\end{equation}
    The value $\omega_{\rm Bh}$ is called angular velocity of rotation
of the black hole.
    For the uniformly rotating coordinate system
    \begin{equation}
\Omega_{(0,1,2)} \to \Omega , \ \ r \to \infty.
\label{OmVrb}
\end{equation}
   So there is an analogy between event horizon of the black hole and
the space infinity in the uniformly rotating coordinate system.
    In its vicinity all physical bodies are rotating with constant
angular velocity.
    It is evident that the coordinate time achieving the event horizon
of the black hole or space infinity in rotating coordinates is infinite.

    Let us study the problem of maximally possible values of radial velocities
of particles in the coordinates of metric~(\ref{gik}).
    Note that due to nonradial motion of massless particles in ergosphere
their radial velocity can have different values (zero for circular orbits).
    As it is known close to the horizon of black holes the coordinate radial
velocity $d r / dt$ approaches zero even in radial falling on
the Schwarzschild black hole.
    This is formulated  as the sentence that the time of falling on the black
hole from the point of view of the observer on the space infinity is infinite.

    Let us find limitations on possible values of the radial velocities of
particles in the metric~(\ref{gik}) from the condition
    \begin{equation}
g_{00} c^2 + 2 c g_{0 \varphi} \frac{d \varphi}{d t} +
g_{\varphi \varphi}\left( \frac{d \varphi}{d t} \right)^2  +
g_{rr} \left( \frac{d r}{d t} \right)^2 +
g_{\theta \theta} \left( \frac{d \theta}{d t} \right)^2\ge 0.
\label{radv}
\end{equation}
    Taking into account that the sum of first three terms in~(\ref{radv})
has the maximal value at
    \begin{equation}
\frac{d \varphi }{ d t} = \frac{\Omega_2 + \Omega_1}{2} =
\frac{ c\, g_{0 \varphi} }{ - g_{\varphi \varphi}},
\label{OmMar}
\end{equation}
    one obtains
    \begin{equation}
\left( \frac{d r }{ d t} \right)^2 \le \frac{g_{\varphi \varphi}}{g_{rr}}
\frac{(\Omega_2 - \Omega_1)^2}{4} = \frac{g_{0 \varphi}^{\,2} -
g_{00} g_{\varphi \varphi}}{g_{rr} g_{\varphi \varphi}}\, c^2.
\label{RadMax}
\end{equation}
    The maximal value $ |d r / d t| $ corresponding to the equality
in~(\ref{RadMax}) corresponds to $\theta = {\rm const}$ and to the angular
velocity~(\ref{OmMar}).
    Note that the angular velocity~(\ref{OmMar}) corresponds to zero
projection of the angular momentum $p_\varphi =0 $ following
from~(\ref{Lpdef}) for the metric~(\ref{gik}).


    For the rotating coordinate system the formula~(\ref{RadMax})
gives trivial result $ |d r / d t| \le c$.
    For the Kerr metric~(\ref{Kerr}) outside the event horizon $r > r_H$
one obtains
    \begin{equation}
\left( \frac{d r }{ d t} \right)^2 \le \frac{\Delta^2}{\Sigma^2} .
\label{RadMaKerr}
\end{equation}
    The representation $\Delta (r) = (r - r_H)(r-r_C)$ leads to the result
that for any moving particle (not only for the free falling)
the coordinate time of achieving the horizon goes to infinity
as $\log (r -r_H)$ for $a<M$ and as $\sim 1/(r -r_H)$ for the extremal
case when $a=M$.

    For particles moving on geodesics one obtains from
Eqs.~(\ref{geodKerr1})--(\ref{geodR})
    \begin{equation}
\left( \frac{d r }{ d t} \right)^2 =
\frac{ \Delta^2 }{ \Sigma^2 } \left[ 1 - \frac{ \Delta }{ \left( \Sigma^2 E -
2 M r a J \right)^2 } \left( \frac{\rho^4 J^2}{\sin^2 \! \theta } +
\Sigma^2 \left( m^2 \rho^2 + \Theta \right) \right) \right] .
\label{RadKerr}
\end{equation}
    So the maximal radial velocity $ |v_r|_{\rm max} = \Delta / \Sigma$
is obtained by massless particles with zero projection of the angular
momentum moving with fixed value of the angle $\theta$ ($\Theta=0$):
    \begin{equation}
\left| \frac{d r }{ d t} \right|_{\rm max} = \frac{\Delta}{ \Sigma}, \ \
{\rm if } \ \ m=0, \ \ J=0, \ \ Q = - a^2 E^2 \cos^2\! \theta.
\label{RadvMa}
\end{equation}

\section{The energy and the momentum in the rotating coordinate system}
\label{secEPVr}

    Energy-momentum vector of the free particle can be obtained
from~(\ref{v6d}) going to the rotating coordinate system
    \begin{equation}    \label{v10}
p^{i} = \frac{ \partial x^i }{ \partial x^{\prime\,k} } p^{\prime\,k}
= \left( \frac{E'}{c}, \ p^{\prime\,r}, \ p^{\prime \varphi}
+ \Omega \frac{E'}{c^2}, \ p^{\prime z} \right),
\end{equation}
    \begin{equation}    \label{v10kk}
p_{i} = \left( \frac{E' + \Omega L'_z}{c},
\ - p^{\prime\,r}, \ - L'_z, \ - p^{\prime z} \right).
\end{equation}
    So in uniformly rotating coordinate system the energy (see~(\ref{dE1}))
and the angular momentum projection are equal to
    \begin{equation}    \label{v10v}
E = E' + \Omega L'_z,  \ \ \ \ L_z = L'_z.
\end{equation}
    The energy of the particle in the rotating coordinate system is
different from that in nonrotating system and can be negative!

    Note that the relativistic formula~(\ref{v10v}) coincides with
the nonrelativistic one presented in the book~\cite{LL_I}.

    One can ask the question: what is the reason for the change of the
expression for the energy in rotating reference frame?
    It seems that the energy defined by~(\ref{NEnergyEgen}) being invariant
is the same in different coordinates?
    However to define the energy in the rotating reference frame one must use
the Killing vector describing translations in time in this reference frame
$\zeta =(1,0,0,0)$.
    Its coordinates in nonrotating frame are $\zeta =(1,0,-\Omega,0)$.
        This Killing vector does not coincide with the Killing
vector $\zeta' =(1,0,0,0)$ describing translations in time in nonrotating frame.
    In rotating frame its coordinates are $\zeta' =(1,0,\Omega,0)$.
        Vector $\zeta$ is the linear combination of different Killing vectors
in nonrotating frame and it can't be obtained by coordinate transformation
from $\zeta'$.
        So defining the energy through translations in time in new reference frame
one obtains new expression for its value.
    Note that the Killing vector of translations in time in rotating system
$\zeta =(1,0,0,0)$ out of the static limit becomes spacelike (as in the case
in Kerr metric) and this explains existence of particles with negative
energy in rotating coordinates.

    The law of conservation of the energy-momentum in interaction of
two particles $(1 , 2 \to 3,4)$ is evidently valid for new values
introduced in rotating coordinate system:
    \begin{equation}    \label{stsn}
\left\{
\begin{array}{l}
E_1 + E_2 = E_3 + E_4 , \\
L_1 + L_2 = L_3 + L_4
\end{array}
\right.
\ \ \Leftrightarrow \ \
\left\{
\begin{array}{l}
E'_1 + E'_2 = E'_3 + E'_4 , \\
L'_1 + L'_2 = L'_3 + L'_4 .
\end{array}
\right.
\end{equation}

    In rotating coordinates the equations of motion of particles with
mass $m$ have the form
    \begin{equation}
c^2 \frac{d t}{d \lambda } = E - \Omega L_z, \ \ \
\frac{d z}{d \lambda } = p^z = {\rm const} ,
\label{uv1}
\end{equation}
    \begin{equation}
r^2 \frac{d \varphi}{d \lambda } = \left( 1 - \frac{\Omega^2 r^2}{c^2}
\right) L_z + \frac{\Omega r^2}{c^2} E,
\label{uv2}
\end{equation}
    \begin{equation}
c^2 \left( \frac{d r}{d \lambda } \right)^2 = R_f, \ \ \ R_f =
(E - \Omega L_z)^2 - m^2c^4 - (p^z c)^2 - \frac{L^2_z c^2}{r^2} .
\label{uv3}
\end{equation}
    The conditions $dt / d \lambda >0$, $R\ge 0$ give the following limitations
on possible values of the energy and momentum.

\begin{figure}[th]
\centering
  \includegraphics[width=7cm]{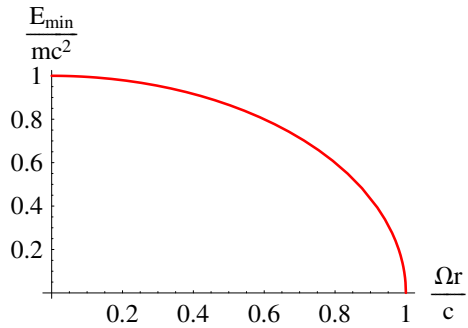}
\caption{The lower boundary of the energy $E_{\rm min}$ in rotating
coordinate system.}
\label{FigE}
\end{figure}

    If $r< c / \Omega $:
    \begin{equation}
E \ge E_{\rm min}, \ \ \
E_{\rm min} = m c^2 \sqrt{1 - \frac{\Omega^2 r^2}{c^2} } .
\label{vmin}
\end{equation}
    The graph of minimal values of the energy is given on Fig.~\ref{FigE}.
    Note that in this region for any particles $E>0$.
        The value $E=0$ is possible only if $m, p^z, L_z=0$, which physically
has no sense because then all energy-momentum components are zero.

    For fixed energy $E$ and the momentum component $p^z$
the following values of the angular momentum are possible
    \begin{equation}
r < \frac{c}{\Omega} \ \ \Rightarrow \ \ E> 0, \ \ \ \
L_- \le L_z \le L_+,
\label{v1r}
\end{equation}
    where
    \begin{equation}
L_\pm = \frac{r}{c}  \frac{\frac{ - \Omega r}{c} E \pm \sqrt{E^2 - \left(
(c p^z)^2 + m^2 c^4 \right) \left( 1 - \frac{ \Omega^2 r^2}{c^2} \right) } }{
1 - \frac{ \Omega^2 r^2}{c^2} } .
\label{vLpm}
\end{equation}

    On the boundary of the static limit $E\ge 0$,
    \begin{equation}
r = \frac{c}{\Omega} \ \ \Rightarrow \ \ \left[
\begin{array}{l}
E > 0, \ \ \ \ \displaystyle
L_z \le \frac{E^2 - m^2 c^4 - (c p^z)^2 }{2 \Omega E} , \\[11pt]
E=0, \ \ m=0, \ \ p^z =0, \ \ \ \ L_z \le 0.
\end{array} \right.
\label{v2r}
\end{equation}

    Out of the static limit the energy in rotating coordinate system $E$
can have as positive as zero and negative values
    \begin{equation}
r > \frac{c}{\Omega} \ \ \Rightarrow \ \ L_z \le L_+.
\label{v3r}
\end{equation}
    One can compare this with~(\ref{EvErg})--(\ref{lHmdd}) for
the Kerr metric.

    On infinity
    \begin{equation}
r \to \infty \ \ \Rightarrow \ \ L_z \le \frac{E -
\sqrt{m^2 c^4 + (p^z c)^2}}{\Omega}.
\label{v3rbesk}
\end{equation}
    so that due to~(\ref{OmBh}), (\ref{OmVrb}) this is analogous
to~(\ref{JH}) in Kerr metric.

    For given values of momenta $p^z$ and $L_z$ the conditions
$dt / d \lambda >0$, $R\ge 0$ lead to limitations on the values of the energy
of particles moving along geodesics~(\ref{uv1})--(\ref{uv3})
    \begin{equation}
E \ge \Omega L_z + \sqrt{ L_z^2 \frac{c^2}{r^2} + (p^z c)^2 + m^2 c^4}.
\label{vrsEVnHS}
\end{equation}
    As for the formula~(\ref{EVnHS}) in case of the Kerr metric one can
conclude that negative energy values are possible only for negative values of
the angular momentum projection $L_z$ out of the static limit $r > c/\Omega$.

    The negative value of the angular momentum due to $L'_z=L_z$ means
the nonzero velocity of particles in Minkowski space in the system at rest.
    Such particles are moving in the coordinate system at rest along lines
in the region $r > c/\Omega$.
    So {\it geodesics of particles with negative energies in the rotating
coordinate system originate and terminate on infinity}\/ analogously to
the event horizon of the Kerr metric in Boyer–Lindquist
coordinates~\cite{GribPavlovVert}.
    They evidently have infinite number of rotations in rotating coordinate
system in direction opposite to the rotation of the system itself.
    This also has the analogy in Kerr metric in Boyer-Lindquist
coordinates~\cite{Chandrasekhar}.

\subsection{Properties of  motion of particles with zero energy  in
rotating coordinate system}
\label{secE0}

    Zero energy in rotating coordinate system is possible for $r \ge c/\Omega$.
    From~(\ref{v2r}), (\ref{v3r}) it is seen that the angular momentum of
the particle with zero energy $L_z \le 0$.
    The angular velocity of particles with zero energy is
    \begin{equation}
\omega_0 = \Omega \left( 1 - \frac{c^2}{\Omega^2 r^2} \right)
\label{omE0}
\end{equation}
    and it does not depend on the angular momentum (i.e on $L_z$) analogous
to the Kerr metric case~\cite{GribPavlov2016}.

    Same analogy is valid for other properties.
        The radial component of the velocity for $E=0$ on the opposite,
depends on the value of the angular momentum projection
    \begin{equation}
E = 0 \ \ \Rightarrow \ \ \frac{d r}{d t} = \pm c\, \sqrt{ 1 -
\frac{c^2}{r^2 \Omega^2} - \frac{m^2 c^4+ (c p^z)^2}{\Omega^2 L_z^2}}
\label{vrt}
\end{equation}
    and is changing for given $r$ (and $m \ne 0$) from zero
    \begin{equation}
E = 0, \ \ \frac{d r}{d t} =0 \ \ \Rightarrow \ \
L_z = - \sqrt{ \frac{m^2 c^4+ (c p^z)^2}{ \Omega^2 - \frac{c^2}{r^2} }}
\label{vrt0}
\end{equation}
    to
    \begin{equation}
\frac{d r}{d t} = \pm c\, \sqrt{ 1 - \frac{c^2}{r^2 \Omega^2}}, \ \
{\rm if} \ \ E = 0, \ \ p^z = {\rm const}, \ \ L_z \to - \infty.
\label{vrtpr}
\end{equation}

        For $E=0$ only particles with $m=0$ and $p^z=0$ achieve
the boundary of the static limit.
    In motionless Minkowski space coordinate system such photons are tangent
to the circle $r = c/\Omega$.
    Let's note that the equation of their trajectory
    \begin{equation}
\varphi (r) - \varphi \left( \frac{c}{\Omega} \right) = \pm \left[
\arcsin \left( \frac{c}{r \Omega} \right)  - \frac{\pi}{2}
+ \sqrt{\frac{r^2 \Omega^2}{c^2} -1 }\right],
\label{TrE0}
\end{equation}
    under replacing
    \begin{equation}
\frac{c}{r \Omega} \to \frac{r c^2}{G M} -1,
\end{equation}
    precisely coincides with the equation of the trajectory of
photons with zero energy (see~\cite{GribPavlov2016}) moving in
the equatorial plane of extremely rotating ($a=G M /c $) Kerr black hole.

    On the boundary of the static limit the angular velocity of particles
with zero energy is equal to zero.
    For $r \to \infty$ the angular velocity of any particle in rotating
coordinate system is going to~$\Omega$ which follows
from~(\ref{uv1}), (\ref{uv2}).

\subsection{The energy of collisions}
\label{Stolk}

    There are many analogies between motion in uniformly rotating coordinate
system and Kerr metric.
    However differently from the Kerr metric the rotating coordinate system
is considered by us for the Minkowski space.
    So physical effects typical for the curved space-time with strong
gravitational field must not be present in this case.

    Let us show this on the example of
the Ba\~{n}ados-Silk-West effect~\cite{BanadosSilkWest09} of the unbounded
growth of the energy of collisions in the center of mass system for motion
in the field of rotating black hole.
    One can find the energy in the center mass system $E_{\rm c.m.}$ of
two colliding particles with rest masses~$m_1$ and $m_2$ taking the square of
the formula
    \begin{equation} \label{SCM}
\left( \frac{E_{\rm c.m.}}{c}, 0\,,0\,,0\, \right) = p^{\,i}_{(1)} + p^{\,i}_{(2)},
\end{equation}
    where $p^{\,i}_{(n)}$ are 4-momenta of particles $(n=1,2)$.
    Due to $p^{\,i}_{(n)} p_{(n)i}= m_n^2$ one has
    \begin{equation} \label{SCM2af}
E_{\rm c.m.}^{\,2} = m_1^2 c^4 + m_2^2 c^4 + 2 p^{\,i}_{(1)} p_{(2)i} c^2 .
\end{equation}
    For free falling particles with energies $E_1$ and $E_2$ and angular
momenta $L_{z 1}$, $L_{z 2}$ from the equations of geodesics in rotating
coordinate system in Minkowski space one obtains
    \begin{eqnarray}
E_{\rm c.m.}^{\,2} &=& m_1^2 c^4 + m_2^2 c^4 +
2 \biggl[ (E_1 - \Omega L_{z 1} ) (E_2 - \Omega L_{z 2} )
\nonumber \\[4mm]
&& -\, \frac{c^2}{r^2} L_{z 1} L_{z 2} - p^z_1 p^z_2 c^2 -
\sqrt{R_{f1} R_{f2}} \biggr].
\label{Col3}
\end{eqnarray}
    However the expression for the energy of collisions in the center of mass
system is relativistic invariant and it can be written through 3-momenta of
particles $\mathbf{p}_n$ in Cartesian coordinates of Minkowski space
    \begin{equation} \label{SCM3i}
E_{\rm c.m.}^{\,2} = m_1^2 c^4 + m_2^2 c^4 + 2 (E_1 E_2 -
c^2 \mathbf{p}_{1} \mathbf{p}_{2}) .
\end{equation}
    So the energy of collisions in rotating coordinate system does not depend
on the point of collision which is different from the Kerr metric where such
dependence is present and is leading to the unbounded growth of the energy of
collision at some point.

\section{On the Penrose process in rotating coordinate system}
\label{secPP}

    The existence of states with negative energy in the ergosphere of
the rotating black hole was used by R.\,Penrose~\cite{Penrose69,Penrose6902}
to consider the process of getting energy from the rotating black hole.
    If the particle in the ergosphere decays on two particles so that
the energy of one of these is negative then the energy of the other
particle must be larger than the energy of the decaying particle.
    If the second particle is going to space infinity then the energy is taken
from the black hole.
    In rotating coordinate system out of the static limit the energy also
can be negative.
    Can one get the energy in the decay of the body on two bodies out of
the static limit?

    Let the massive particle at the rest system at
the point $ x_0 = \alpha c / \Omega $  (for example $\alpha = 5$) decays on
two particles with the velocity $v' = \beta c $ (for example $\beta = 3/5$)
moving along the vertical in opposite directions.
    Then at the rest frame
    \begin{equation} \label{Pen1}
E' = mc^2 , \ \ \ E'_1 = E'_2 = \frac{E'}{2}, \ \ \
L'_{z 1} = - L'_{z 2} = \frac{E'}{2\,c^2} v' r.
\end{equation}
    In rotating coordinate system with angular velocity $\Omega$
    \begin{equation} \label{Pen2}
E = E' , \ \ \
E_i = E'_i \pm \Omega L_{z i} = E'_i \left( 1 \pm \frac{\Omega v' r}{c^2}
\right)= \frac{E}{2} ( 1 \pm \alpha \beta ).
\end{equation}
    For taken $ \alpha = 5$, $\beta = 3/5$:
    \begin{equation} \label{Pen3}
E_1 = 2 E, \ \ \ E_2 = - E .
\end{equation}
    So the energy of the first resulting particles is larger than the energy
of the initial particle.
    To realize the analogue of the Penrose process one must get the particle
with the energy $E_1 > E$ at the point of observation.
    However if this observation is at the point zero of coordinate system then
the angular momentum $L_{z i}$ must be equal to zero and the energies of
particles in rotating coordinate system $E_i = E'_i + \Omega L_{z i}$
at the moment of observation occur to be the same $E_i=E'/2$.

\begin{figure}
\centering
  \includegraphics[width=7cm]{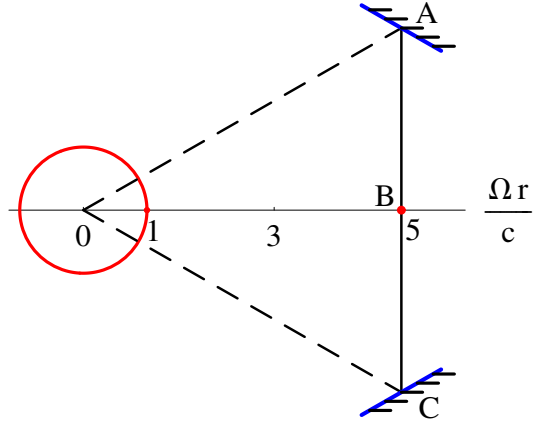}
\caption{The decay of particles at the point B and reflection at screens A, C.}
\label{FigPenroseRot}
\end{figure}

    For example these final particles can be directed without change of
the energy at rest system by use of absolutely elastic collisions at
points $(x_0, \pm y) $  with mirrors standing under the angles
$\frac{3}{4} \pi + \frac{1}{2} \tan^{-1}( \frac{y}{x_0}) $,
$\frac{\pi}{4} - \frac{1}{2} \tan^{-1}( \frac{y}{x_0}) $, as it is shown on
Fig.~\ref{FigPenroseRot}.
    In the rotating coordinates this looks like collision with moving screens.
      Particle with negative energy collides with the screen moving to it and
gets the energy.
    Particle with positive energy collides with the screen going from it and
gives energy to it.
    In the result both energies become the same equal to one half of
the energy of the decaying particle.

    If one makes the observation not at the zero point (the center of
the Earth) the angular momentum can be nonzero and the energy in rotating
coordinates can be different from that at rest.
    Then the Penrose effect can be observable.
    It occurs due to motion of the observer relative to the rest frame
to or out of the moving particle.
    The order of the effect is defined by the linear velocity of the observer
which is proportional to the distance from the rotation axis
    \begin{equation} \label{Pen4}
E = E' + \Omega L_z = E' \left( 1 + \frac{r^2}{c^2} \Omega \omega \right),
\end{equation}
    where $\omega$ is the angular velocity of the observed particle.
        In case of observation of relativistic particles (for example in cosmic
rays) its angular velocity can get values $\omega = \pm c /r$ and the energy
in rotating coordinates
    \begin{equation} \label{Pen5}
E = E' \left( 1 \pm \frac{r}{c} \Omega \right).
\end{equation}
    In observation of such particles on the equator of
the Earth $r \approx 6378$\,km  the additional term in the energy
is of the order $1.55 \cdot 10^{-6}$.

\section{Conclusion}
\label{Concl}

    As it is seen from the obtained results the description of motion of
particles in Boyer-Lindquist coordinates for the Kerr metric and
the description of particle motion in rotating coordinates in flat
space-time have many general features.
    In the rotating system and in Kerr metric there are regions where
particles cannot be at rest: the ergosphere in Kerr metric and the region
out of the surface $r = c /\Omega$ in rotating coordinate system.
    States of particles with negative and zero energies are possible in
these regions.

    There are analogous inequalities on values of the energy and of
the angular momentum projection for motion of particles divided by the static
limit (see (\ref{EvErg})--(\ref{EVnHS}) for the Kerr black hole
and (\ref{v1r})--(\ref{vrsEVnHS}) for the rotating system).
    The angular velocity of particles in these regions has always the same
direction and is limited between the maximal and minimal values.

    Inside the ergosphere and in the region $r > c /\Omega$ in rotating
system the value of the particle angular velocity defines the sign of
the particle energy (see Fig.~\ref{VrOmega}).

    There is analogy between the event horizon of the Kerr black hole
and radial space infinity $( r \to \infty)$ of the rotating coordinate system.

    In both cases movement to them takes infinite coordinate time and
the angular velocity is going to a definite limit: to the angular velocity
of the rotation of the black hole or to the angular velocity of
the coordinate system.
    In both cases one needs infinite number of rotations to achieve
either the horizon or $( r \to \infty)$.
    Geodesics of particles with zero and negative energies originate
and terminate in $r=r_H$ in Kerr metric or in radial infinity
$r \to \infty$ in rotating coordinate system.

    However one must note that there is no full analogy between motion
in Kerr metric and that in rotating coordinates in flat space-time.
    There is no Ba\~{n}ados-Silk-West effect~\cite{BanadosSilkWest09}
of the growth of the energy in the center of mass frame for two colliding
particles present in Kerr metric but absent in rotating coordinates in
Minkowski space.
    The Penrose effect formally is present in both cases but it has
observable consequences only far from the origin point of the rotating
coordinate system.

    All results concerning Minkowski space-time in rotating coordinates
are new.
    They show what the observer using these coordinates will see.
    We hope that these results will be found interesting by
any researcher of General Relativity.

\begin{acknowledgements}
    This work was supported by the Russian Foundation for Basic Research,
grant No. 15-02-06818-a and by the Russian Government Program of
Competitive Growth of Kazan Federal University.
\end{acknowledgements}



\begin{thebibliography}{99}
\label{Ref}

\bibitem{Ehrenfest1909}
Ehrenfest, P.:
Gleichf\"{o}rmige Rotation starrer K\"{o}rper und Relativit\"{a}tstheorie.
\href{http://dx.doi.org/10.1007/978-94-017-0528-8_1}
{Phys. Z. {\bf 10}, 918 (1909)}
[English transl.
``Uniform Rotation of Rigid Bodies and the Theory of Relativity''
in~\cite{EhrenfestVrash}, p.\,3]

\bibitem{EhrenfestVrash}
Rizzi, G., Ruggiero,  M.L. (eds.)
\href{http://dx.doi.org/10.1007/978-94-017-0528-8}
{Relativity in Rotating Frames}.
Kluwer Academic Publ., Boston (2004)

\bibitem{LL_II}
Landau, L.D., Lifshitz, E.M.: The Classical Theory of Fields.
Pergamon Press, Oxford (1994)

\bibitem{Fok}
Fock, V.: Theory of Space, Time and Gravitation.
Pergamon Press, Oxford (1964)

\bibitem{Kerr63}
Kerr, R.P.:
Gravitational field of a spinning mass as an example of algebraically
special metrics.
\href{http://dx.doi.org/10.1103/PhysRevLett.11.237}
{Phys. Rev. Lett. {\bf 11}, 237--238 (1963)}

\bibitem{BoyerLindquist67}
Boyer, R.H., Lindquist, R.W.:
Maximal analytic extension of the Kerr metric.
\href{http://dx.doi.org/10.1063/1.1705193}
{J. Math. Phys. {\bf 8}, 265--281 (1967)}

\bibitem{MTW}
Misner, C.W., Thorne, K.S., Wheeler, J.A.: Gravitation.
Freeman, San Francisco (1973)

\bibitem{NovikovFrolov}
Novikov, I.D., Frolov, V.P.:
Physics of Black Holes [in Russian]. Nauka, Moscow (1986);
Frolov, V.P., Novikov, I.D.:
Black Hole Physics: Basic Concepts and New Developments.
Kluwer Acad. Publ., Dordrecht, (1998)

\bibitem{Chandrasekhar}
Chandrasekhar, S.: The Mathematical Theory of Black Holes.
Oxford University Press, Oxford (1983)

\bibitem{LL_I}
Landau, L.D., Lifshitz, E.M.: Mechanics.
Pergamon Press, Oxford (1976).

\bibitem{GribPavlov2013}
Grib, A.A., Pavlov, Yu.V.:
On the energy of particle collisions in the ergosphere of the rotating
black holes.
\href{http://dx.doi.org/10.1209/0295-5075/101/20004}
{Europhys. Lett. {\bf 101}, 20004 (2013)}

\bibitem{GribPavlov2012}
Grib, A.A., Pavlov, Yu.V.:
Collision energy of particles in the ergosphere of rotating black holes.
\href{http://dx.doi.org/10.1007/s11232-013-0075-4}
{Theor. Math. Phys. {\bf 176}, 881--887 (2013)}

\bibitem{Zaslavskii13c}
Zaslavskii, O.B.:
Acceleration of particles as a universal property of ergosphere.
\href{http://dx.doi.org/10.1142/S0217732313500375}
{Mod. Phys. Lett. A {\bf 28}, 1350037 (2013)}

\bibitem{LPPT}
Lightman, A.P., Press, W.H., Price, R.H., Teukolsky, S.A.:
Problem book in relativity and gravitation.
Princeton University Press, Princeton, New Jersey (1975)

\bibitem{GribPavlovVert}
Grib, A.A., Pavlov, Yu.V., Vertogradov, V.D.:
Geodesics with negative energy in the ergosphere of rotating black holes.
\href{http://dx.doi.org/10.1142/S0217732314501107}
{Mod. Phys. Lett. A {\bf 29}, 1450110 (2014)}

\bibitem{GribPavlov2016}
Grib, A.A., Pavlov, Yu.V.:
Black holes and particles with zero or negative energy.
\href{http://dx.doi.org/10.1134/S0040577917020088}
{Theor. Math. Phys. {\bf 190}, 268--278 (2017)};
\href{http://arxiv.org/abs/1601.02592}
{arXiv:1601.02592}

\bibitem{BanadosSilkWest09}
Ba\~{n}ados, M., Silk, J., West, S.M.:
Kerr black holes as particle accelerators to arbitrarily high energy.
\href{http://dx.doi.org/10.1103/PhysRevLett.103.111102}
{Phys. Rev. Lett. {\bf 103}, 111102 (2009)}

\bibitem{Penrose69}
Penrose, R.:
Gravitational collapse: The role of General Relativity.
{Rivista Nuovo Cimento {\bf I}, Num. Spec., 252--276 (1969)}

\bibitem{Penrose6902}
Penrose, R.:
Gravitational collapse: The role of General Relativity.
\href{http://dx.doi.org/10.1023/A:1016578408204}
{Gen. Relativ. Gravit. {\bf 34}, 1141--1165 (2002)}

\end{thebibliography}
\end{document}